# Observation of the n($^3$He,t)p Reaction by Detection of Far-Ultraviolet Radiation

Alan K. Thompson[1], Michael A. Coplan[2,4], John W. Cooper[2], Patrick Hughes[3], Robert E. Vest[4] and Charles W. Clark[4,5]

[1]Ionizing Radiation Division, National Institute of Standards and Technology, Gaithersburg, MD  20899-8461
[2]Institute for Physical Science and Technology, University of Maryland, College Park, MD  20742
[3]Department of Physics, University of Maryland, College Park, MD 20742
[4]Electron and Optical Physics Division, National Institute of Standards and Technology, Gaithersburg, MD  20899-8410
[5]Joint Quantum Institute, National Institute of Standards and Technology and University of Maryland, Gaithersburg, MD  20899-8410

We have detected Lyman alpha radiation as a product of the n($^3$He,t)p nuclear reaction occurring in a cell of $^3$He gas.  The predominant source of this radiation appears to be decay of the 2p state of tritium produced by charge transfer and excitation collisions with the background $^3$He gas.  Under the experimental conditions reported here we find yields of tens of Lyman alpha photons for every neutron reaction.  These results suggest a method of cold neutron detection that is complementary to existing technologies that use proportional counters.  In particular, this approach may provide single neutron sensitivity with wide dynamic range capability, and a class of neutron detectors that are compact and operate at relatively low voltages.

## 1. Introduction

Nuclear physics and atomic, molecular and optical (AMO) physics take place on disparate energy scales, with nuclear excitations having characteristic energies of 1 MeV ($10^6$ electron volts) while the energies encountered in atomic electron excitation are from 10 eV to 1000 eV.  This is one of the reasons that nuclear and AMO physics have largely developed as independent disciplines, with different experimental and theoretical techniques.  There has always been some interplay between them, such as the determination of nuclear moments and charge distributions from high-resolution atomic spectroscopy [1].  The present investigation is at the boundary between these two disciplines.  It is concerned with atomic electron excitation during the course of the low-energy n($^3$He,t)p reaction.  We have found that this reaction generates a quantity of Lyman alpha radiation that is easily detectable in $^3$He gas targets at pressures from 5.3 kPa to 106 kPa (40 Torr to 800 Torr), an observation which, to our knowledge, has not been previously made. This radiation may serve as a useful alternative signature for precise neutron dosimetry, perhaps leading to single-neutron detection capabilities or as a vehicle for compact neutron detectors that can operate over a wide dynamic range without high voltages.

The overwhelming majority of neutron detectors in use today are either $BF_3$ or $^3He$ gas proportional tubes. $BF_3$ is a toxic and corrosive gas. While there are many instruments using $BF_3$ in the field, manufacturers are moving away from its use. $^3He$ proportional tube detectors have efficiencies similar to the proposed new technique, but require high voltages (1300 V to 2000 V) to operate, are susceptible to microphonics and have a dead time of approximately 1 μs that limits their maximum counting rate. The tubes also require an ultra-pure quench gas (usually $CO_2$) to achieve a sufficient signal-to-noise ratio. They also suffer from wall effects when particle energy is lost by absorption at the tube walls [2, 3].

There are a small number of detectors using a lithium-doped scintillator (lithium glass, lithium iodide, or lithium-loaded plastic). Their utility is limited by gamma ray backgrounds. Fission chamber detectors that use the reaction of neutrons with isotopes of uranium or plutonium to produce energetic charged particles are used for neutron flux calibration measurements, but are not generally practical for routine measurements because of low sensitivity [3].

The $n(^3He,t)p$ reaction has long been studied in nuclear physics [4], and now is the basis of most thermal neutron detectors used at the National Institute of Standards and Technology (NIST). The small uncertainty in the cross section (0.12%) suggests this reaction as a candidate for the primary standard detector for accurate determination of thermal neutron fluence, but the accuracy of $^3He$ proportional tubes (a few percent) is more than an order of magnitude larger than the uncertainty in the cross section. Improvements in the primary accuracy of neutron fluence measurement would yield improvements in the knowledge of the neutron lifetime, and through that tighten bounds on tests of the Standard Model [5]. The reaction is exothermic at zero incident neutron energy, where it yields (from the nuclear perspective alone), a triton and a proton with combined escape energy of 764 keV [6]. This reaction is between a low-energy neutron and a $^3He$ atom in a $^3He$ gas. From an AMO or few-body physics perspective, this reaction is of interest because the speeds of the reactant triton and proton are comparable to the "speeds" of the atomic electrons in the ground state of He, and thus in the range of the Massey criterion for efficient charge transfer [7]. As will be shown, our experiment indicates that charge transfer plays a dominant role in Lyman alpha production.

Since the interaction between an incident slow neutron and the atomic electrons can be disregarded in the first instance, this reaction can be described from an AMO perspective as nearly an ideal example of high-energy molecular dissociation. Here a TH ($^3H^1H$) diatomic molecule is instantaneously prepared in the united-atom limit [8], receives a large amount of energy and almost no momentum, and then dissociates with 764 keV excess energy. The reaction yields, with unknown branching ratios, hydrogen atoms (H), tritium atoms (T), protons (p), tritons (t), and electrons (e) in a number of final state configurations:

t + p + 2e
T + H

T⁻ + p
t + H⁻
t + H + e
T + p + e

In addition, T and H in the above may be replaced by T* and H* to denote the production of excited atomic states.

Here we offer estimates of the importance of the various final state configurations in our measurements. The reaction leads to tritons and protons, or tritium and hydrogen atoms, at energies of 191 and 573 keV respectively, corresponding to respective speeds of $3.5 \times 10^6$ m•s$^{-1}$ and $1.05 \times 10^7$ m•s$^{-1}$, comparable to the $4.3 \times 10^6$ m•s$^{-1}$ average speed of electrons in the ground state of helium which is obtained from the virial theorem [9]. In an ambient environment of $^3$He gas, Lyman alpha radiation can be produced by the following reactions: H(2p) or T(2p) from the initial reaction; higher excited states of H and T from the initial reaction followed by radiative or collisional relaxation to 2p states; ground states of H and T from the initial reaction followed by subsequent collisions with $^3$He; and direct production of protons and tritons followed by charge exchange collisions with $^3$He, leading to both ground and excited states of H and T, that then undergo subsequent collisions. These mechanisms are all potential sources of Lyman alpha photons, to which the local $^3$He environment is transparent. We first report the results of observations of Lyman alpha radiation in $^3$He over a range of pressures, and discuss the relative importance of the underlying mechanisms.

## 2. Experiment

The experiment consists of a reaction cell, a photomultiplier-detector of Lyman alpha radiation, processing and recording electronics, and a gas handling system. A diagram of the reaction cell is shown in Fig. 1. The cell is a stainless steel cube with 70 mm vacuum seal surfaces machined into each of the six faces. Collimated neutrons from a beamline at the NIST Center for Neutron Research enter the cell through a silicon window 24 mm in diameter and 0.5 mm thick on the front face. A lithium-impregnated plastic disk at the end of the cube interior opposite the silicon window acts as a beam dump, absorbing all unreacted neutrons. A 25-mm diameter thin-walled tube of magnesium is mounted vertically inside the cube in order to precisely define the field of view of the detector. The tube is transparent to neutrons and opaque to Lyman alpha photons.

Lyman alpha photons produced in the magnesium tube within the reaction cell were detected with a Hamamatsu solar blind R6835 photomultiplier tube in a modified model 658 end-on housing from McPherson Instruments, Inc. The R6835 has a MgF$_2$ window and a CsI photocathode. The nominal quantum efficiency of the photomultiplier tube is 10 % at a wavelength of 120 nm and falls to 1 % at 180 nm. The typical gain is $6 \times 10^5$ at 2200 V. The housing was mounted on the top of the reaction cell behind a 29.2 mm diameter MgF$_2$ window. A differentially pumped volume between this window and the photomultiplier prevented helium from coming into contact with the photomultiplier tube which could render it inoperative over time. The photomultiplier was operated in the photon counting mode with its output directly connected to a spectroscopy amplifier whose output drove the input of a multichannel analyzer.

The gas handling system consisted of a stainless steel manifold with connections through high-vacuum valves to the reaction cell, the volume between the $MgF_2$ window and the photomultiplier tube, a cylinder containing $^3He$, and an oil-free scroll pump with a base pressure of 67 Pa (50 mTorr). The pressure in the reaction cell was monitored with a digital pressure indicator (Omega DPI 705) that covered the range up to 200 kPa (1500 Torr ) and requires no correction for gas type. Absolute pressure could be measured to 13 Pa (0.1 Torr ). The incident neutron flux was measured with a fission chamber to a combined relative standard uncertainty of 5 % as detailed in Sec. 2.2.

**2.1 Detector Calibration**

The Lyman alpha photomultiplier detector was calibrated absolutely against a NIST calibrated photodiode at the Far UV Calibration Facility [10] at NIST. Continuous Lyman alpha radiation was generated by a hydrogen discharge and narrow band monochromator. To approximate the conditions of the experiment as closely as possible, the calibration configuration included the $MgF_2$ window and the differentially pumped volume, as well as the photomultiplier tube and reaction cell. The calibration procedure consisted of first measuring the incident Lyman alpha flux with a calibrated silicon photodiode reference detector located in the photon beam path just upstream from the reaction cell. Subsequently, the detector was completely withdrawn to allow the beam to enter the reaction cell. The current produced by the calibrated detector was recorded and converted to incident photon flux from knowledge of the absolute efficiency of the photodiode.

Because the photomultiplier tube has a gain of $10^5$ to $10^6$, the radiation from the monochromator was attenuated to avoid over loading the photomultiplier tube. This was accomplished with a stack of four interference filters from Acton Optics and Coatings that were manufactured in a single deposition run. The filters were located at the center of the reaction cell. As measured by the manufacturer, each filter has a nominal maximum transmission of 8.6% at 119.2 nm and full width half maximum (FWHM) bandwidth of 8.7 nm. The nominal transmission for Lyman alpha radiation (121.6 nm) is 7.32%; the calculated transmission of the stack of four filters is $2.87 \times 10^{-5}$. The transmission of the filter stack was determined experimentally by measuring the ratio of the photocurrents from two Si photodiodes – the calibrated photodiode and a similar photodiode mounted in place of the photomultiplier tube. The photocurrent ratio was measured with the filters installed and again with the filters removed. The filter stack transmission is the photocurrent ratio with the filters installed divided by that without the filters. This ratio-of-ratios technique eliminates any dependence of the filter stack transmission on the absolute value of detector calibration and any requirements on the long-term stability of the source. In order to have sufficient signal from the photodiode behind the filter stack, the monochromator slit widths were increased to raise the radiation intensity by a factor of approximately $10^5$ over that used in the photomultiplier calibration. The measured transmission of the filter stack is $(2.24 \pm 0.03) \times 10^{-5}$, which differs from the nominal transmission value by about 20%, indicating a 6% difference per filter. The measured transmission value was used in subsequent photomultiplier calibration measurements. The same technique was used to determine the $MgF_2$ window transmission of $0.543 \pm 0.003$.

The use of strongly attenuating filters makes the control of scattered light imperative. Even very small amounts of scattered photons from the monochromator exit slit or other scattering surfaces can exceed the flux transmitted through the filters. The response of the photomultiplier to these scattered photons can lead to a systematic error of efficiency by orders of magnitude. To shield the active area of the photomultiplier from scattered photons, an aluminum tube was installed between the rear of the filter stack and the $MgF_2$ window in front of the photomultiplier. This tube effectively limited the field of view of the photomultiplier to the filter stack. With this baffling in place, scattered light was eliminated.

Operation of the photomultiplier tube in the photon-counting mode required that the output charge pulses corresponding to photoelectrons from the photocathode be converted to voltage pulses that could be analyzed and recorded by a multichannel analyser (MCA). This was accomplished with a shaping amplifier. Pulse height distributions from the MCA were transferred to a computer for analysis. The photon flux incident on the $MgF_2$ window was determined from the measured photon flux at the reference detector and the transmission of the filter stack. Pulses from the photomultiplier were recorded and the output pulse rate calculated from the MCA pulse height spectrum and acquisition live time. The photomultiplier response rate is determined by subtracting a dark count rate from the measured output rate. Both the output and dark count rates are determined by integrating the respective MCA spectra over all channels and dividing by the acquisition live time. The efficiency is the ratio of the response count rate to the incident photon rate. Typical photodiode currents were of the order of $10^{-11}$A corresponding to an incident Lyman alpha photon flux on the order of $10^8$ $s^{-1}$ at the reference detector. The filter stack reduced this rate to several thousand per second at the $MgF_2$ window. The transmission of the $MgF_2$ window and the efficiency of the photomultiplier resulted in an output pulse rate of approximately an order of magnitude less.

The collection efficiency of the window-photomultiplier system for Lyman alpha photons (the ratio of observed pulses to incident photons) was measured as $0.0764 \pm 0.0076$. Given the window transmission value reported above, the quantum efficiency of the photomultiplier tube is $0.141 \pm 0.014$. This value compares well with the nominal value supplied by the manufacturer.

### 2.2 Measurements

The cell depicted in Fig. 1 was mounted on a neutron beamline at the NIST Center for Neutron Research, with a 4 mm diameter neutron beam as indicated in the figure. The neutron beam flux was established absolutely with a calibrated fission detector, and was measured to be $2.31 \times 10^4$ $s^{-1}$ with a combined relative standard uncertainty of 5 %. The wavelength of the neutrons was (0.496 ± 0.001) nm [11]. For neutrons with this wavelength, the cross section [6] for the reaction n($^3$He,p)t is (1.471 ± 0.002) x $10^{-20}$ cm$^2$.

Experimental data were taken by filling the evacuated reaction cell with $^3$He (Spectra Gases, 99.9999 % chemical purity and 99.95 % isotopic purity), and recording the total number of photomultiplier tube pulses registered by the multichannel analyzer during 1000 s acquisition intervals. The pressure of $^3$He in the chamber was recorded before and after data acquisition. Normally these two pressure readings differed by at most 13 Pa (0.1 Torr). After the acquisition period, the data were recorded and the reaction cell pressure was changed by either letting in additional $^3$He from the manifold or evacuating $^3$He through the manifold. Data were then acquired for another 1000 s period at the new pressure. This procedure was repeated for $^3$He pressures from approximately 5.3 kPa to 106 kPa (40 Torr to 800 Torr).

Because the nominal quantum efficiency of the photomultiplier tube is 10 % at 120 nm and falls to 1 % at 180 nm, a series of experiments were performed to determine the degree to which the observed photon signal was indeed due to Lyman alpha radiation. These experiments consisted of recording the photomultiplier tube signal as described above and then repeating the measurements with a narrow band Lyman alpha filter (used in the detector calibration) in front of the photomultiplier tube. In all cases the ratio of the signals with and without the filter corresponded to the measured transmission of the filter. We take this as important evidence that the observed signal is Lyman alpha radiation.

The uncertainty in the measurements of photon intensity as a function of $^3$He pressure for a constant neutron intensity is limited by the background signal from the photomultiplier tube. This background has a number of sources: dark counts originating from within the photomultiplier tube, background radiation generated from radiation sources external to the experiment and background radiation generated within the reaction cell from the incident neutrons. We evaluated the background by recording photomultiplier pulses with the reaction cell evacuated and also with the cell filled with different pressures of $^4$He. The results of the background experiments gave an average background rate of (9.4 ± 0.5) s$^{-1}$. The background was checked periodically during the course of the experiment.

The experimental results are shown in Fig. 2. The data points, through which the solid line has been drawn, represent the photomultiplier count rate corrected for the background and divided by the measured efficiency. The error bars include the statistical uncertainties of signal and background rates and the 10 % uncertainty in the photomultiplier tube calibration. The broken curve is the calculated neutron reaction rate within the magnesium tube. The calculation takes into account the n($^3$He,t)p cross section, the measured incident neutron flux, and the geometry of the reaction cell. To understand the nature of the results it is necessary to recognize that there is attenuation of the incident neutron beam in the region between the silicon entrance window and the 25 mm diameter magnesium tube. The maximum neutron reaction rate in the magnesium tube occurs at about 60 kPa (450 Torr) while the maximum observed photon rate occurs at about 87 kPa (650 Torr). In order to estimate the Lyman alpha production rates we divide the observed photon rates by the ratio of the solid angle subtended by the photocathode (0.054 sr) to $4\pi$ sr. This assumed that the photon emission can be approximated as originating from a point at the center of the reaction cell. These Lyman alpha production rates divided by the corresponding neutron reaction rate give the number of photons produced per reacted neutron. Fig. 3 shows the Lyman alpha photon yield per reacted neutron. If the production of Lyman alpha photons were a result of the n($^3$He,p)t reaction alone, the number of photons per reacted neutron should be constant, independent of $^3$He pressure. Instead, the data in Fig. 3 fall on a curve. We take this as strong evidence that Lyman alpha photons are produced in reactions that depend on $^3$He pressure occurring after the primary n($^3$He,p)t reaction. This will be discussed in more detail in the Sec. 3. At 93 kPa (700 Torr), 46 photons are produced for every neutron reacting with $^3$He. This high yield of photons is the main result of this investigation and can be the basis of a neutron detector that can complement existing detectors.

## 3. Modeling

In order to understand the processes that produce Lyman alpha radiation as a result of the n($^3$He,t)p reaction we consider (1) formation of t and p in the initial nuclear reaction and (2) production of Lyman alpha from the reaction products of (1).

The modeling of (1) is rather straightforward and requires only a knowledge of the neutron flux, the $^3$He pressure, the n($^3$He,t)p cross section, and the experiment geometry. For the 0.496 nm cold neutron beam the only reaction that has a large cross section is neutron absorption due to the n($^3$He,t)p reaction. For example, using the cross section of $1.47 \times 10^{-20}$ cm$^2$, and a chamber filled with $^3$He at 101 kPa (760 Torr) the mean free path for neutrons will be 27 mm. and 60% of the neutrons incident on the 25 mm magnesium tube will be absorbed within the tube.

The results quoted above indicate that direct production of Lyman alpha photons in the n($^3$He,p)t reaction is not the source of the observed signal. Tritons and protons are ejected by the initial reaction and emerge in opposite directions with equal momenta so that the energy of the proton will be three times that of the triton; there is no preferred orientation of their trajectories with respect to the neutron beam since the incident neutron has essentially zero momentum. Assuming all of the reaction energy is given to the ejected particles (the binding energy of the two electrons in $^3$He is 79 eV) the triton and proton energies will be 191 and 573 keV respectively. The charged particles will be produced in a 4 mm diameter cylinder with length set by the 25 mm diameter of the magnesium cylinder.

The assumption that the initial reaction produces only tritons and protons greatly simplifies the modeling of the processes that are likely to produce Lyman alpha radiation. First, if only ions are formed there must be a charge exchange process occurring, either resulting in or followed by excitation. Second, from a standpoint of the atomic processes taking place there is no distinction between tritons and protons and the reactions for tritons can be thought of as simply excitation and/or charge exchange between helium atoms and protons and hydrogen atoms allowing for the difference in masses between protons and tritons. There has been a great deal of experimental and theoretical work in the last sixty years on the reactions of protons with helium and hydrogen. Consequently, accurate cross sections [12] are available for most of the processes producing Lyman alpha radiation in our apparatus.

With the above assumptions, the only ways that Lyman alpha radiation can be produced are:

1. Direct charge transfer
   p + He → H(2p) + He$^+$ → H(1s) + Lyman alpha

2. Charge transfer followed by excitation
   p + He → H(1s) + He$^+$
   H(1s) + He → H(2p) + He → H(1s) + Lyman alpha

In 1 and 2 we have written reactions only for protons (p) and hydrogen atoms (H). The same reactions occur for tritium (T) and tritons (t) taking into consideration the difference in mass between protons and tritons. In identifying these processes we have ignored the possibility of excitation to higher excited states of H followed by cascade to H(2p) and subsequent production of Lyman alpha radiation. This is a reasonable assumption since measurements of the cross sections for charge exchange accompanied by excitation generally found that cascade effects were of the order of 5% at most [11]. We have also considered the possibility that the observed radiation is a result of transitions from the n=4 to n=2 level of $^3$He$^+$, which occurs at 121.5 nm and thus cannot be easily resolved from the Lyman alpha radiation at 121.6 nm. The effect of these transitions on the measurement of 2p excitation cross sections in charge changing collisions of protons and helium has been considered by Hippler *et. al.* [13]. They find that for incident energies lower than approximately 300 keV the effect of these transitions on the observed Lyman alpha radiation may be ignored. One of the key results of earlier work on the interaction of protons with gases is the realization that at kinetic energies below 1 MeV for pressures near 101 kPa (760 Torr), protons continuously associate and dissociate with electrons from the gas as they lose energy. In order to model the production of Lyman alpha we make the following assumptions:

1. During the slowing down of t or p within our apparatus there are a number of charge changing collisions that result in the atoms being uncharged for a fraction of the time required for the slowing down process.

2. During the slowing down process there is a possibility of Lyman alpha production from both reactions 1 and 2.

We adopt a simple model of the slowing-down process by assuming that all energy loss is due to charge changing collisions. Following the work of Allison [14], the time τ spent in a neutral state of a beam of particles slowing down from initial energy $E_0$ to a final energy $E$ will be:

$$\tau = \frac{p_0 T}{p T_0} \left| \int_{E_0}^{E} \frac{\sigma_{10}(E)}{\sigma_{01}(E) + \sigma_{10}(E)} \frac{dE}{\varepsilon(E) V(E)} \right|$$

Here p and T are the pressure and temperature in the reaction cell, $p_0$ and $T_0$ are the standard pressure (101.3 kPa) and temperature (273 K), $\sigma_{01}(E)$ and $\sigma_{10}(E)$ are the cross sections for charge exchange from neutral (0) to ionized (1), $\varepsilon(E)$ is the stopping power in units of keV•cm$^{-1}$ and $V(E)$ is the velocity of the fast moving particles in cm•s$^{-1}$. The time spent in a singly ionized state can be obtained simply by interchanging $\sigma_{01}(E)$ and $\sigma_{10}(E)$. The charge changing cross sections for helium have been measured [15,16] and are given in [13]. The stopping power for protons in helium is well known [17] and the stopping power for tritons can be inferred from the proton data.

During the time the particle is neutral, production of Lyman alpha photons can occur via reaction 2 above and via reaction 1 when the particle is ionized. The cross sections for Lyman alpha formation by protons and hydrogen atoms have been measured [12] and the analogous cross sections for tritons can be obtained from these data. Within our model there are four possible sources of Lyman alpha radiation, namely radiation produced by hydrogen atoms or protons or by tritium or tritons. In principle it is possible to calculate the ratios of the various processes; we have not done so but can identify the dominant processes from an analysis of the cross sections.

First, consider production of Lyman alpha radiation from reaction 1. Within our model a proton colliding with helium can capture an electron into the H ground state or into an excited state, of which the 2p state is the most probable. At 100 keV the cross sections for these two processes are 3.4 x 10$^{-17}$ cm$^2$ and 3.9 x 10$^{-19}$ cm$^2$. The cross section for Lyman alpha production by reaction 1 peaks with proton energy of 20 keV but it is still 50 times smaller than the charge exchange cross section leading to ground state atoms. From this we conclude that electron capture into excited states is not the most important process. The same is not true for reaction 2. While the cross section for Lyman alpha production by H is approximately an order of magnitude smaller than the charge exchange cross section at energies above 100 keV the same is not true at lower energies. The cross section for Lyman alpha production via reaction 2 peaks with proton energy of 400 eV and is seven times larger than the charge exchange cross section.

From the above we conclude that most of the radiation we observe comes from excitation of neutral atoms of H and T after they have been slowed to below 1 keV. Lyman alpha production at higher kinetic energies is not probable since the lifetime of the 2p level which must be excited to produce it is far longer that the average time between collisions.

It also seems that the radiation from T is probably the dominant process. The ranges of p and t in helium produced by the neutron absorption have already been determined [2]. The results, 51.6 mm for protons and 14.6 mm for tritons at standard pressure (101.3 kPa) and temperature (273 K), indicate that at the lower pressures used here the initially formed protons will have been lost to the walls of the reaction chamber before reaching the low energies necessary to produce Lyman alpha photons in large quantities.

## 4. Conclusions

We have observed that the reaction of cold neutrons with $^3$He produces Lyman alpha photons with a yield of 46 photons per reacted neutron at a $^3$He pressure of 93 kPa (700 Torr). The production of the photons is primarily a result of collisions of the initial reaction products, tritons and protons with the background $^3$He gas. Since $^3$He is transparent to 121.6 nm Lyman alpha radiation, there is no attenuation of the photon signal in the reaction cell. The cross sections for all the relevant collisions have either been measured or calculated, and it is possible to quantitatively model the photon production process. This will be the subject of a future publication.

The high yield of Lyman alpha photons from the reaction of neutrons with $^3$He can be the basis of a sensitive thermal neutron detector that has a wide dynamic range and advantages over existing detectors. As discussed in Sec. 1, existing detectors have significant limitations which this new technique may be able to circumvent.

In the experiments described here, no attempt has been made to optimize the Lyman alpha yield. It is clear that larger signal rates can be obtained by modifying the geometry of the reaction cell and through the use of efficient radiation detectors with large acceptance angles. By adding other moderator gases to $^3$He, it is also possible to increase the yield of Lyman alpha radiation from the energetic H atoms. We believe that the observations reported here can lead to a new class of cold neutron detectors that combine sensitive, single neutron detection with wide dynamic range and a relatively simple, compact and robust package.

**Acknowledgements**

This work was motivated by suggestions made by Anthony Leggett. James McGuire provided helpful advice on atomic collision physics. We thank the NIST Center for Neutron Research for access to the NG6 neutron beam line. Mr. Hongdian Yang assisted with detector calibration, experimental measurements and data analysis.

**Note:** Certain commercial equipment, instruments, or materials are identified in this paper to foster understanding. Such identification does not imply recommendation or endorsement by the National Institute of Standards and Technology, nor does it imply that the materials or equipment identified are necessarily the best available for the purpose.

**About the Authors:** Alan K. Thompson is a physicist in the Ionizing Radiation Division of the NIST Physics Laboratory. Michael A. Coplan is Professor and Director of the Chemical Physics Graduate Program at the University of Maryland, College Park. John W. Cooper is Faculty Research Associate at the Institute for Physical Science and Technology, University of Maryland, College Park, and a former Physics Laboratory Editor of the NIST Journal of Research. Patrick Hughes is a post-doctoral researcher in the Chemical Physics Graduate Program at the University of Maryland, College Park. Robert E. Vest is a physicist in the Electron and Optical Physics Division of the NIST Physics Laboratory. Charles W. Clark is Chief of the Electron and Optical Physics

Division of the NIST Physics Laboratory and a Fellow of the Joint Quantum Institute, National Institute of Standards and Technology and University of Maryland.

**Figure Captions**

**Figure 1.** Scale drawing of the experimental reaction cell. All dimensions in mm. The 4 mm diameter neutron beam enters the cell through a silicon window attached to the left face of the cell. A 25 mm diameter magnesium tube in the cell limits the radiation that can reach the photomultiplier detector mounted on the top of the cell behind a $MgF_2$ exit window. An evacuated volume between the $MgF_2$ window and the photomultiplier tube eliminates the possibility of $^3$He coming in contact with the face of the tube and diffusing into it. Not shown are the gas, vacuum, and pressure meter connections to the cell.

**Figure 2.** Observed Lyman alpha photon rate as a function of $^3$He pressure (solid line, solid circles) and calculated reacted neutrons (broken line) as a function of $^3$He pressure. The measured efficiency of the photomultiplier tube has been used to calculate the photon rate from the signal rate. The calculated neutron reaction rate is based on the $n(^3He,p)t$ cross section, measured incident neutron rate, and reaction cell geometry. The uncertainties in the measured data points reflect the statistical uncertainties of the measurements as well as the uncertainty in the calibration of the photomultiplier tube detector. The solid line through the data points is provided to guide the eye.

**Figure 3.** Observed Lyman alpha photons per neutron reaction as a function of $^3$He pressure. In calculating the photon production rate, the solid angle of the detector was taken into consideration. The uncertainties in the measured data points reflect the statistical uncertainties of the measurements, the uncertainty in the calibration of the photomultiplier tube detector, but not the uncertainty in the collection efficiency of the detector which we estimate to be of order 20 %. The solid line through the data points is provided to guide the eye.

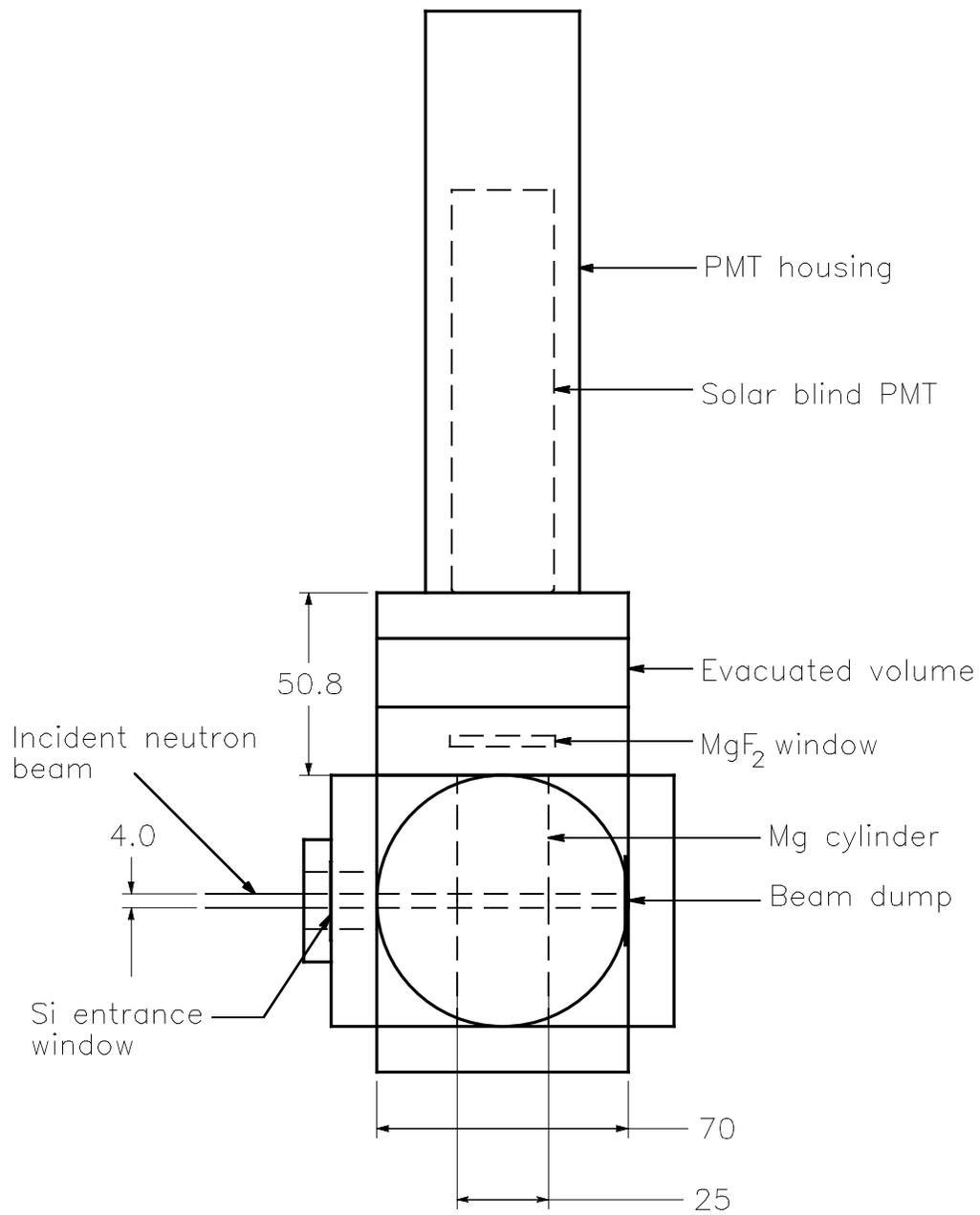

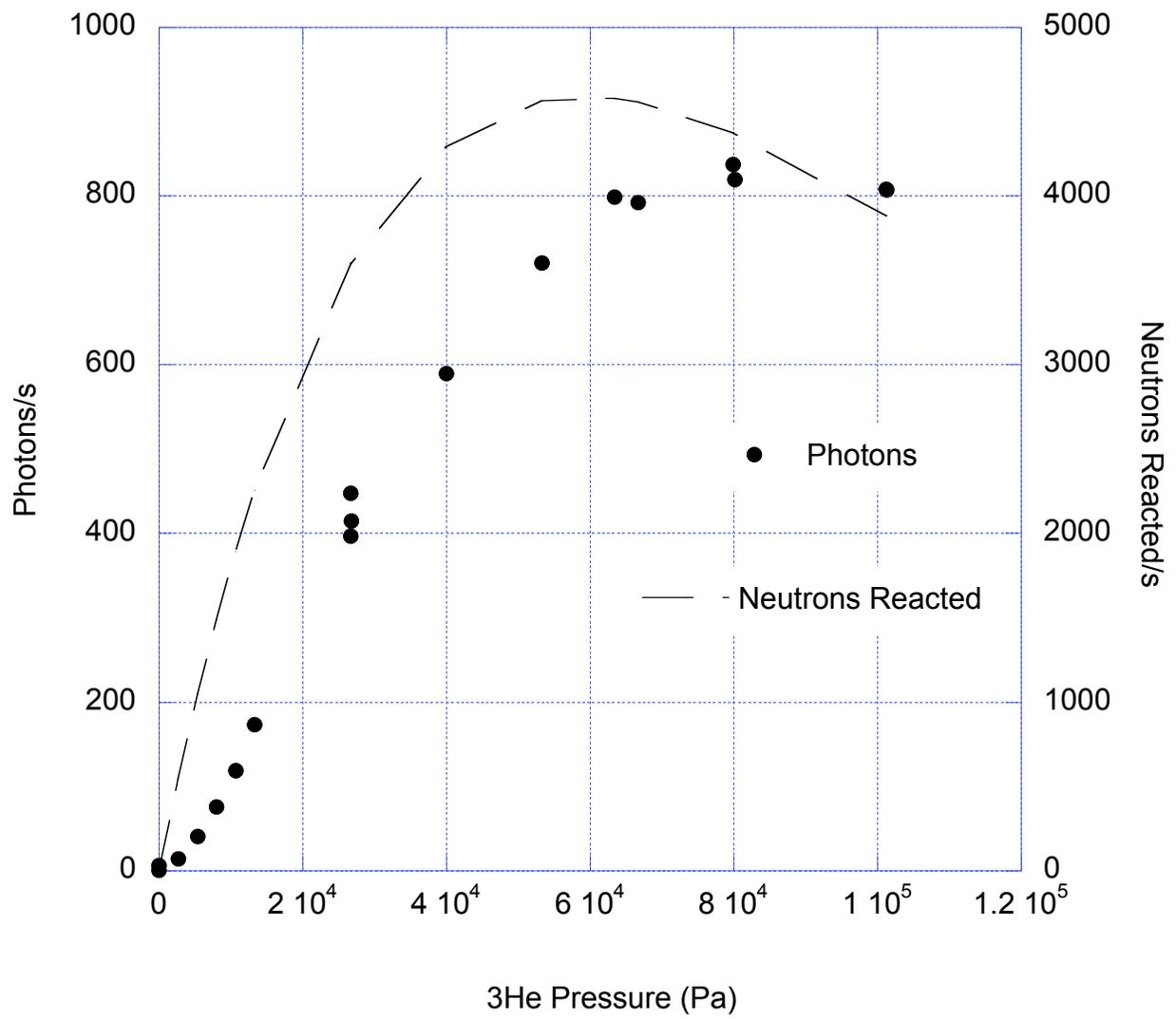

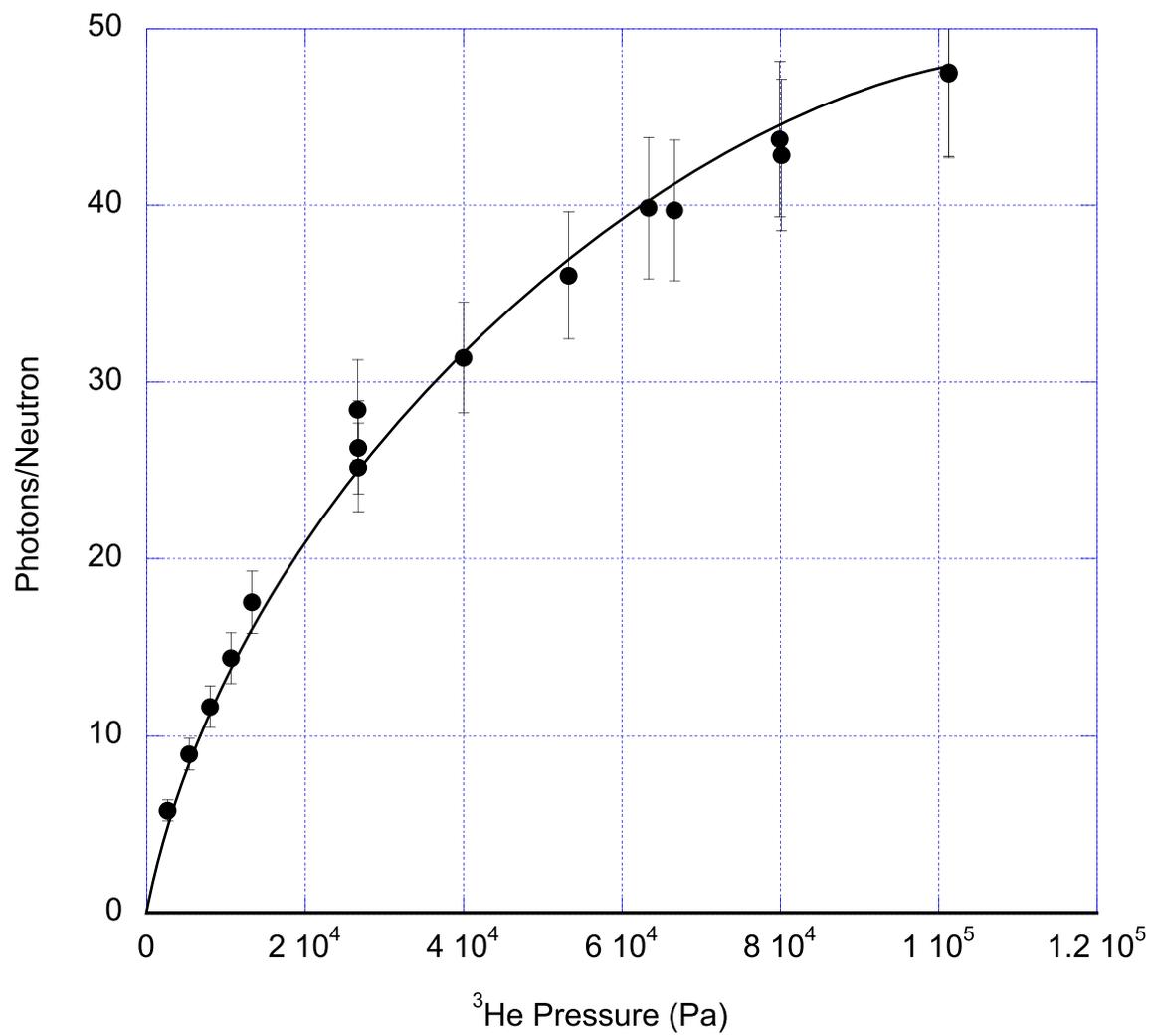